\documentclass[10pt,conference]{IEEEtran}

\usepackage[letterpaper,left=0.65in,right=0.65in,top=0.75in,bottom=1.08in]{geometry}

\usepackage{amsmath,amssymb,amsfonts,bm}

\usepackage{graphicx}
\usepackage{subcaption}
\usepackage[export]{adjustbox}
\usepackage{epstopdf}

\usepackage{array}
\usepackage{multirow}
\usepackage[table]{xcolor}
\usepackage[flushleft]{threeparttable}

\usepackage{algorithm}
\usepackage{algpseudocode}
\usepackage{booktabs}

\usepackage{cite}
\usepackage{enumitem}
\usepackage{comment}
\usepackage{orcidlink}
\usepackage{changepage}

\usepackage{tikz}
\usetikzlibrary{positioning,shapes.geometric,arrows.meta}

\usepackage{mathptmx}

\usepackage{pifont}

\definecolor{lightGray}{RGB}{217,217,217}

\newcolumntype{L}[1]{>{\raggedright\arraybackslash}m{#1}}
\newcolumntype{C}[1]{>{\centering\arraybackslash}m{#1}}
\newcolumntype{R}[1]{>{\raggedleft\arraybackslash}m{#1}}

\begin{document}

\title{Uncertainty-Aware and Explainable Ensemble Deep Learning Framework for Multi-Class Skin Lesion Classification}

\author{
    \IEEEauthorblockN{Rofiqul Islam\IEEEauthorrefmark{1}, Lilatul Ferdouse\IEEEauthorrefmark{1}} 
    \IEEEauthorblockA{\IEEEauthorrefmark{1}Department of Computer Science and Physics, Wilfrid Laurier University, Waterloo, ON, Canada\\
    Email: \{isla5070@mylaurier.ca, lferdouse@wlu.ca\}}
}


\maketitle

\begin{abstract}
Skin cancer diagnosis from dermoscopic images remains challenging due to high intra-class variability, inter-class similarity, class imbalance, and the limited interpretability of deep learning models. This paper proposes an uncertainty-aware and explainable deep learning framework for multi-class skin lesion classification.
The framework combines a vision transformer model (MaxViT-Tiny) with CNN-based models (ConvNeXt-Tiny and EfficientNetV2-B0) through deep ensemble learning. Monte Carlo (MC) Dropout estimates predictive uncertainty and identifies unreliable predictions, while Grad-CAM++, an explainable AI (XAI) technique, provides visual explanations by highlighting lesion regions that influence model decisions. Evaluated on the HAM10000 dataset, the framework achieves $96\%$ accuracy and $99\%$ ROC-AUC under uncertainty-aware filtering (entropy $<1.0$, confidence $\geq0.7$), with macro-average precision, recall, and F1-score of $94\%$, $95\%$, and $95\%$, respectively, and $96\%$ weighted-average scores across all three metrics. The results demonstrate accurate, interpretable, and uncertainty-aware skin lesion classification for trustworthy computer-aided diagnosis.

\end{abstract}

\begin{IEEEkeywords}
Skin lesion classification, deep ensemble learning, uncertainty estimation, explainable artificial intelligence (XAI), Monte Carlo dropout, Grad-CAM++.
\end{IEEEkeywords}

\section{Introduction}

\label{Intro}

Skin cancer is one of the most common cancers worldwide and represents a significant public health challenge. It includes multiple clinically distinct skin lesion categories that vary in appearance, biological characteristics, and treatment approaches. Early and accurate diagnosis is crucial for improving patient outcomes and reducing disease progression, particularly for aggressive forms such as melanoma~\cite{zhou2025global}. Dermoscopy is a widely adopted non-invasive imaging technique that enables dermatologists to examine skin lesions in greater detail and improves diagnostic accuracy compared with conventional visual examination~\cite{kittler2002diagnostic}. However, accurate interpretation of dermoscopic images requires substantial clinical expertise, and diagnostic decisions may vary among dermatologists due to differences in experience and subjective assessment. Furthermore, the limited availability of specialized dermatologists in many healthcare settings emphasizes the need for automated, reliable, and interpretable computer-aided diagnostic systems.

Recent advances in deep learning have significantly improved medical image analysis by enabling automatic extraction of discriminative features from complex images~\cite{litjens2017survey}. Convolutional Neural Networks (CNNs) have demonstrated strong capability in learning local visual patterns and hierarchical feature representations~\cite{Chen2024ARO}, while Vision Transformers (ViTs) employ self-attention mechanisms to capture long-range dependencies and global contextual information from images~\cite{dosovitskiy2020image}. Hybrid CNN-transformer architectures combine the strengths of convolutional feature extraction and transformer-based global representation learning, achieving promising performance in medical image classification tasks~\cite{zhang2021TransFuseFT}. In dermoscopic image analysis, these models can capture important lesion characteristics, including texture, pigmentation, shape, border irregularities, and structural patterns, which are valuable for distinguishing different skin lesion categories. However, many existing studies~\cite{ozdemir2025robust,lilhore2025skinehdlf,mavaddati2025skin,abdulredah2025towards,shakya2025comprehensive} primarily emphasize improving classification performance, while uncertainty estimation and model interpretability remain relatively less explored, despite their importance for reliable clinical deployment.

Despite these advances, accurate classification of multiple skin lesion types remains challenging. Dermoscopic datasets often suffer from class imbalance, where common lesion categories substantially outnumber rare but clinically important classes. Other challenges include high intra-class variability, strong inter-class visual similarity, image artifacts, and the limited availability of annotated medical data, all of which can affect model performance and generalization~\cite{tschandl2018ham10000}. Moreover, many deep learning models produce deterministic predictions without quantifying predictive uncertainty, which may limit their reliability in clinical decision-making~\cite{gal2016dropout}. In addition, the limited interpretability of deep learning models reduces clinicians' trust and limits their adoption in routine clinical practice~\cite{arrieta2020explainable}.

To overcome these limitations, this work develops an uncertainty-aware and explainable deep learning framework for multi-class skin lesion classification. A deep ensemble of MaxViT-Tiny, ConvNeXt-Tiny, and EfficientNetV2-B0 is combined with Monte Carlo (MC) Dropout to estimate predictive uncertainty via multiple stochastic forward passes, enabling the identification of low-confidence cases for further clinical assessment~\cite{gal2016dropout}. Furthermore, Grad-CAM++ is utilized to visualize the regions contributing to model decisions, enhancing model interpretability and supporting transparent analysis of classification outcomes~\cite{chattopadhay2018grad}.

The motivation of this research is based on both clinical and technical needs. Clinically, an uncertainty-aware and explainable multi-class diagnostic system can support dermatologists by providing reliable predictions, reducing diagnostic variations, and helping detect different types of skin lesions at an early stage, especially in areas with limited healthcare resources. Technically, the integration of deep ensemble learning, uncertainty estimation, explainable AI, and advanced CNN and transformer models aims to improve classification accuracy, robustness, and model transparency. Therefore, this work contributes to the development of reliable and practical AI-based systems for automated skin cancer diagnosis. The main contributions of this paper are summarized as follows:

\begin{itemize}

\item We propose an uncertainty-aware deep ensemble framework that combines a vision transformer model (MaxViT-Tiny) with CNN-based models (ConvNeXt-Tiny and EfficientNetV2-B0) for robust multi-class skin lesion classification.

\item We integrate deep ensemble learning with Monte Carlo  Dropout to estimate predictive uncertainty, enabling reliable and confidence-aware clinical decision support.

\item We employ Grad-CAM++ to generate visual explanations, improving the interpretability and transparency of the classification process.

\item We validate the proposed framework on the HAM10000 dataset, achieving $96\%$ accuracy and $99\%$ ROC-AUC under the uncertainty-aware filtering criterion (entropy $<1.0$ and confidence $\geq0.7$).

\item The proposed uncertainty-aware framework supports clinically relevant operating scenarios by enabling high-coverage large-scale skin lesion assessment and high-confidence point-of-care diagnosis.

\end{itemize}







The organization of this paper is as follows. Section \ref{Intro} presents the clinical motivation, discusses the existing research gaps, and summarizes the main contributions. Section \ref{method} describes the proposed framework, including data preparation, ensemble-based classification, uncertainty estimation, and explainability. Next, Section \ref{evaluation} presents the performance evaluation, including the experimental results, and confusion matrix analysis. Section \ref{clinical} discusses potential clinical deployment scenarios, including large-scale skin lesion assessment with high prediction coverage and point-of-care decision support with high-confidence predictions. Finally, Section \ref{conclu} concludes the paper and outlines future research directions.

\section{Methodology}
\label{method}

\begin{figure}
\centering     
\includegraphics[width=\linewidth, height=7cm]{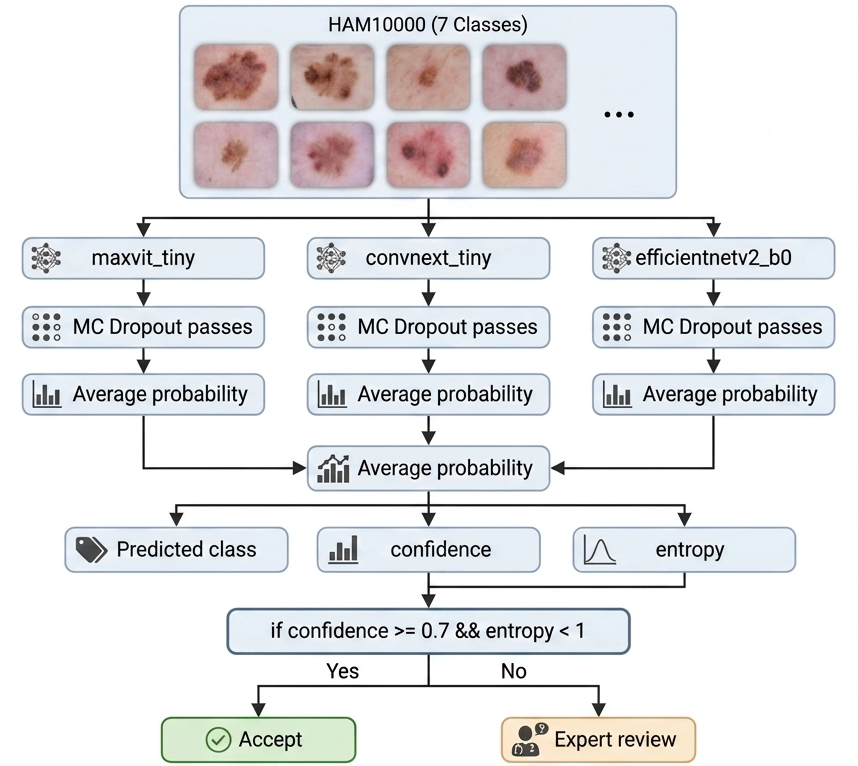}
\caption {\centering  Uncertainty-Aware Ensemble Deep Learning Framework for Multi-Class Skin Lesion Classification.}
\label{fig:main_fig}
\end{figure}

The proposed uncertainty-aware deep learning framework for multi-class skin lesion classification is illustrated in Fig.~\ref{fig:main_fig}. The framework is designed to improve classification reliability by combining deep ensemble learning, uncertainty quantification, and selective prediction. 



          
\subsection{Dataset and Preprocessing}

The proposed framework is evaluated on the HAM10000 dataset~\cite{tschandl2018ham10000}, a widely adopted benchmark for dermoscopic skin lesion classification. The dataset consists of $10,015$ RGB dermoscopic images collected from diverse clinical sources, covering multiple lesion categories with considerable variations in visual appearance and imaging conditions. Each image is expert-annotated and associated with metadata, enabling reproducible evaluation and stratified dataset partitioning.

Introduced as part of the ISIC 2018 challenge~\cite{codella2019skin}, the HAM10000 dataset comprises seven diagnostic classes: benign keratosis-like lesions (bkl), melanocytic nevi (nv), dermatofibroma (df), melanoma (mel), vascular lesions (vasc), basal cell carcinoma (bcc), and actinic keratoses and intraepithelial carcinoma (akiec). These categories exhibit significant variations in color, texture, morphology, and lesion size, presenting challenges for automated dermoscopic image classification. To ensure unbiased evaluation, the dataset is divided into training ($80\%$) and testing ($20\%$) subsets using stratified sampling. This approach preserves the original class distribution across subsets and enables reliable assessment of model performance.

All images were resized to $224 \times 224$ pixels and normalized using the standard ImageNet normalization parameters, with mean values 
$\mu=[0.485, 0.456, 0.406]$ and standard deviation values 
$\sigma=[0.229, 0.224, 0.225]$. During training, data augmentation techniques, including random horizontal and vertical flipping, random rotation up to $25^\circ$, color jittering with brightness, contrast, and saturation variations of 0.2, and random affine transformations with a shear factor of $10$ degrees, are applied to improve model robustness and generalization. To mitigate the impact of class imbalance, a weighted sampling strategy is employed to provide balanced exposure to minority lesion classes during model training.

\subsection{Deep Ensemble Classification}

To improve classification robustness, a deep ensemble of three state-of-the-art architectures is constructed, including MaxViT-Tiny, ConvNeXt-Tiny, and EfficientNetV2-B0. These models provide complementary feature representations by combining convolution-based local feature extraction and transformer-based global dependency modeling.Each pretrained model is initialized using ImageNet weights and fine-tuned for seven-class skin lesion classification. The original classification layer is replaced with a task-specific fully connected layer corresponding to the seven lesion categories. During training, focal loss~\cite{lin2017focal} is employed to mitigate class imbalance by assigning greater emphasis to difficult and minority-class samples. The focal loss is defined as
\begin{equation}
\begin{aligned}
\mathcal{L}_{\mathrm{FL}}
=
-\alpha(1-p_t)^{\gamma}\log(p_t),
\label{eq:focal_loss}
\end{aligned}
\end{equation}
where $p_t$ denotes the predicted probability of the ground-truth class, $\alpha$ is the class-balancing factor, and $\gamma$ is the focusing parameter that reduces the contribution of easily classified samples while emphasizing hard examples.

\subsection{Uncertainty Estimation}

To quantify predictive uncertainty, MC Dropout~\cite{gal2016dropout} is combined with deep ensemble learning~\cite{lakshminarayanan2017simple} to capture both model uncertainty and prediction variability. During inference, dropout layers remain active, and multiple stochastic forward passes are performed for each input image. Specifically, each ensemble model performs $T=20$ stochastic forward passes, where $m$ indexes the ensemble model. The predictive probability of each model is obtained by averaging the stochastic outputs:
\begin{equation}
\begin{aligned}
\bar{p}_{m}(y|x)
=
\frac{1}{T}
\sum_{t=1}^{T}p_{m,t}(y|x),
\end{aligned}
\end{equation}
where $p_{m,t}(y|x)$ represents the predicted probability from the $t$-th stochastic forward pass of the $m$-th ensemble model. The final predictive distribution is obtained by averaging the outputs of all $M=3$ ensemble models:
\begin{equation}
\begin{aligned}
p(y|x)
=
\frac{1}{M}
\sum_{m=1}^{M}\bar{p}_{m}(y|x),
\end{aligned}
\end{equation}
where $M$ denotes the number of ensemble models. The predictive uncertainty is quantified using predictive entropy:
\begin{equation}
\begin{aligned}
\mathcal{H}(p)
=
-\sum_{c=1}^{C}p_c\log(p_c),
\end{aligned}
\end{equation}
where $C=7$ represents the number of lesion classes and $p_c$ denotes the predicted probability of class $c$. Higher entropy values indicate a more uncertain prediction distribution, whereas lower entropy corresponds to confident model predictions.

\subsection{Uncertainty-Aware Selective Prediction}

To improve reliability, a selective prediction strategy is adopted by rejecting uncertain predictions. A prediction is considered reliable when both entropy and confidence satisfy predefined criteria:
\begin{equation}
\begin{aligned}
\mathcal{H}(p)<\tau_e,\quad \max(p_c)\geq\tau_c,
\end{aligned}
\end{equation}
where $\tau_e$ and $\tau_c$ represent the entropy and confidence thresholds, respectively. In this work, $\tau_e=1.0$ and $\tau_c=0.7$ are used to retain high-confidence predictions for evaluation. Samples that do not satisfy these criteria are considered uncertain and can be referred to dermatologists for further clinical assessment, supporting reliable AI-assisted decision-making.

\subsection{Explainability Analysis}

To improve model interpretability, Grad-CAM++ is employed to visualize discriminative image regions influencing the classification decisions. The final feature extraction stage of the MaxViT-Tiny backbone is selected as the target layer for generating class-specific activation maps based on the predicted lesion category. 
The generated heatmaps highlight image regions that have a strong influence on the model's classification decisions, providing visual evidence for the predictions and improving the transparency of the AI-based diagnostic process.

\begin{figure}[!ht]
\centering
\begin{subfigure}[b]{0.48\linewidth}
    \centering
    \includegraphics[width=\linewidth]{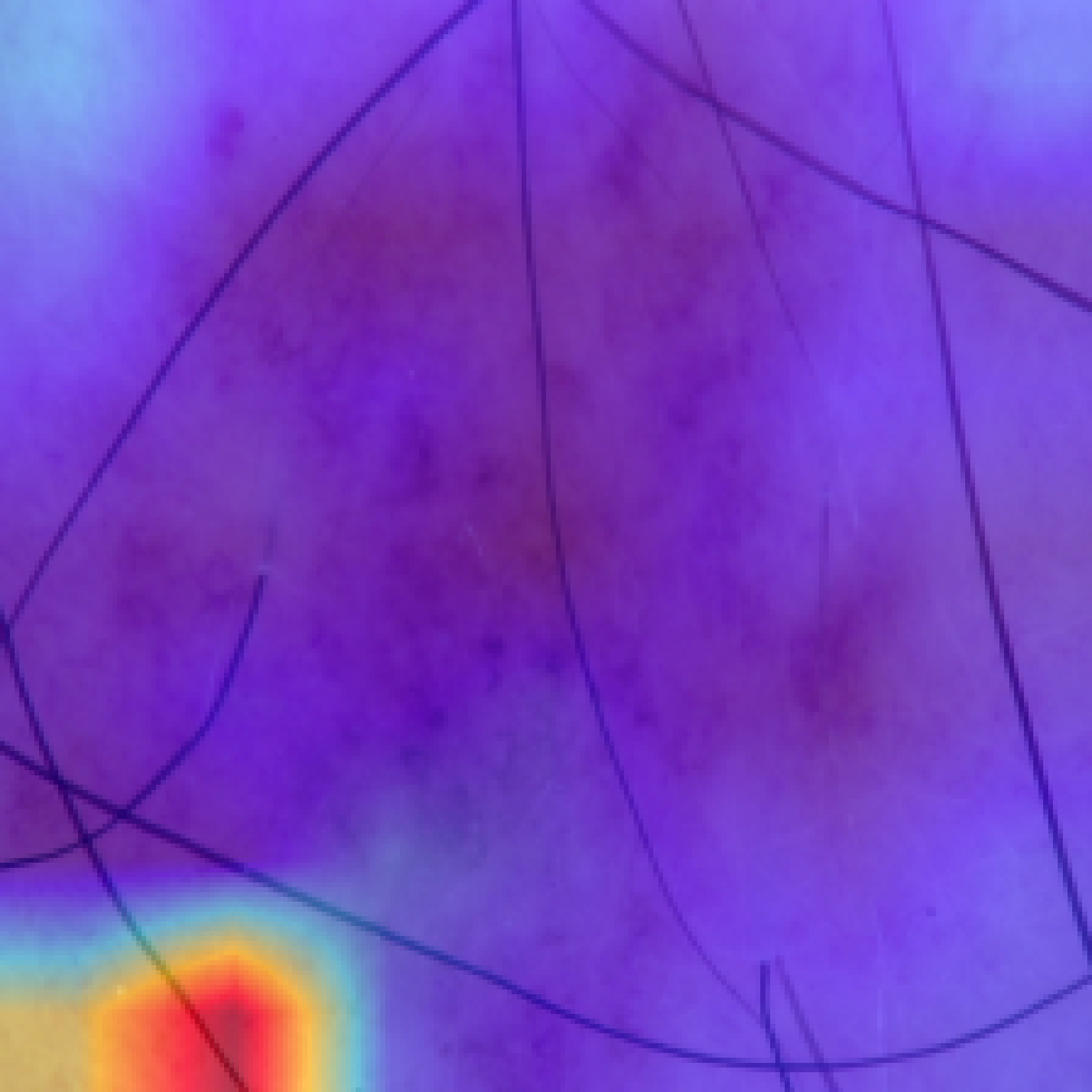}
    \caption{High-confidence nv}
    \label{fig:gradcam_nv}
\end{subfigure}
\hfill
\begin{subfigure}[b]{0.48\linewidth}
    \centering
    \includegraphics[width=\linewidth]{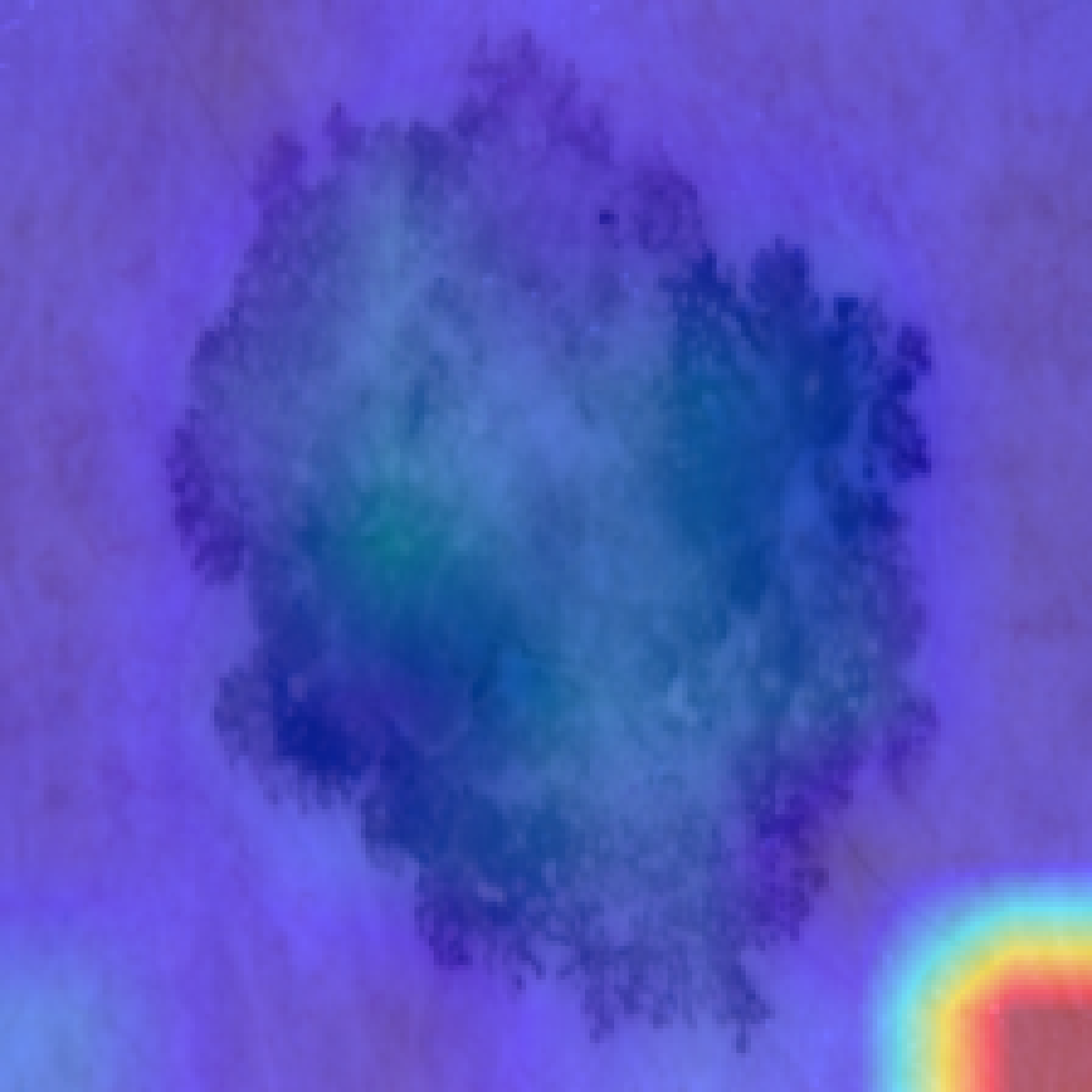}
    \caption{High-confidence mel}
    \label{fig:gradcam_mel}
\end{subfigure}

\vspace{1mm}

\begin{subfigure}[b]{0.48\linewidth}
    \centering
    \includegraphics[width=\linewidth]{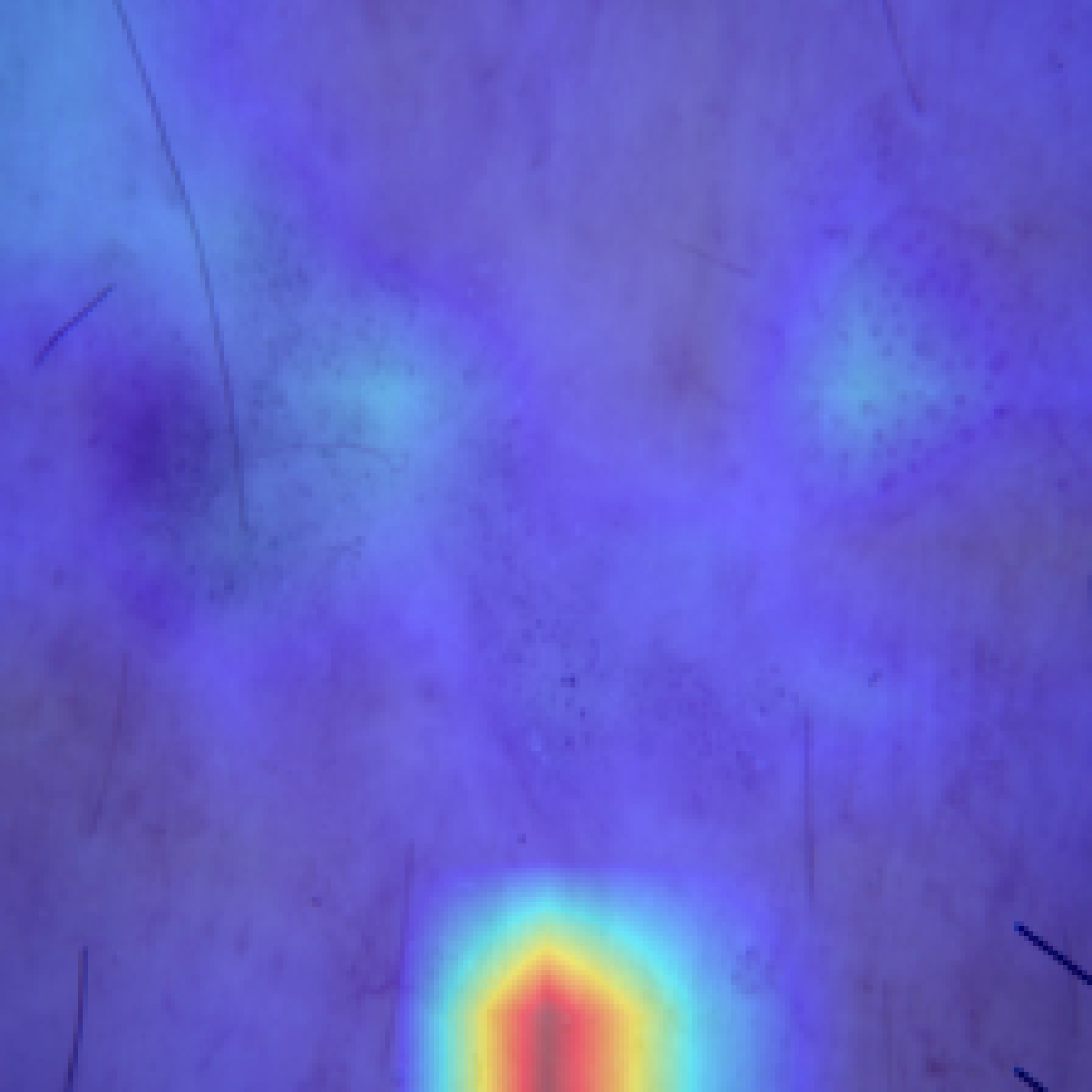}
    \caption{High-confidence akiec}
    \label{fig:gradcam_akiec}
\end{subfigure}
\hfill
\begin{subfigure}[b]{0.48\linewidth}
    \centering
    \includegraphics[width=\linewidth]{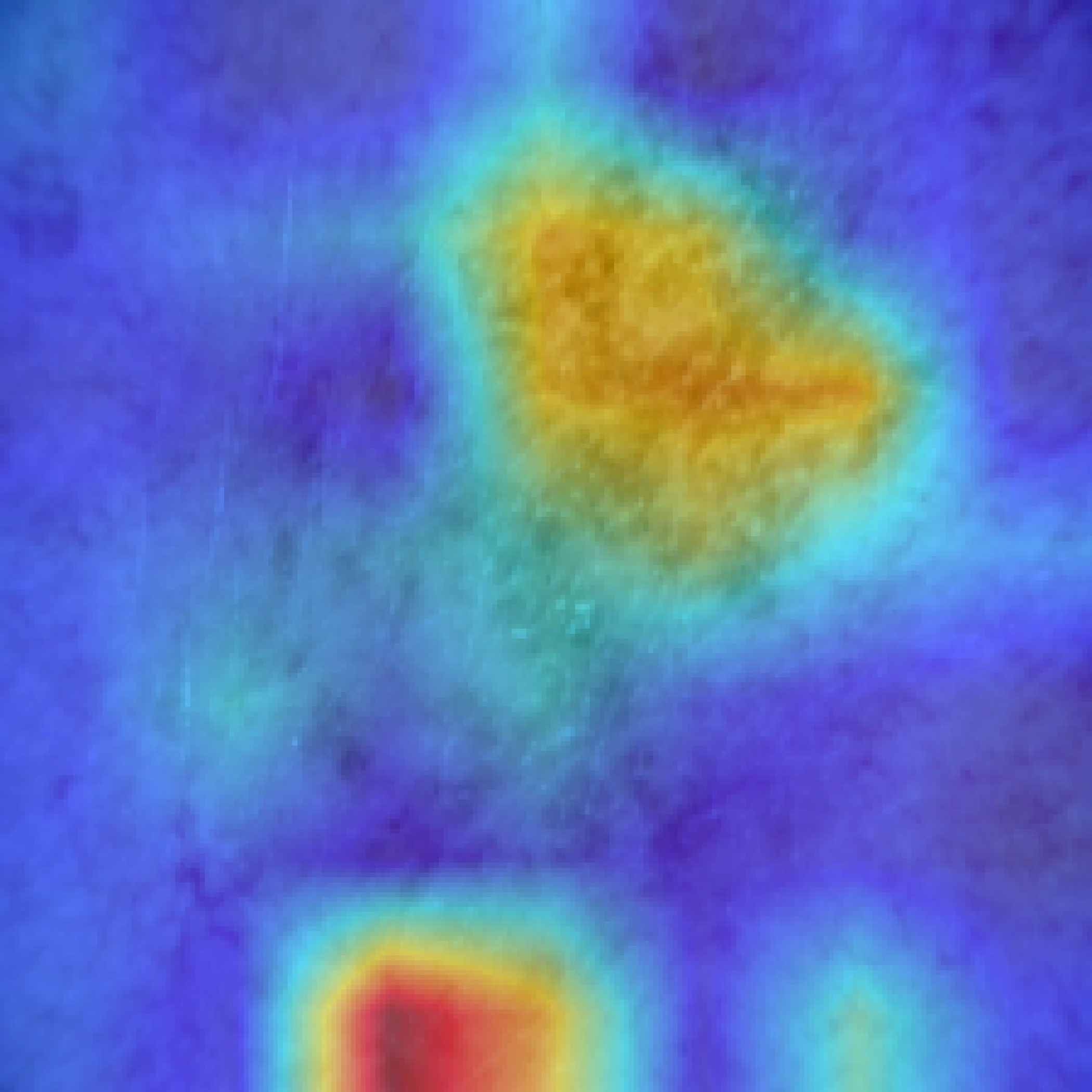}
    \caption{Uncertain bkl}
    \label{fig:gradcam_bkl}
\end{subfigure}

\caption{Grad-CAM++ visualization of representative skin lesion predictions: 
(\subref{fig:gradcam_nv}) correctly classified high-confidence nv, 
(\subref{fig:gradcam_mel}) correctly classified high-confidence mel, 
(\subref{fig:gradcam_akiec}) correctly classified high-confidence akiec, 
and (\subref{fig:gradcam_bkl}) a correctly classified but uncertain bkl prediction identified by high entropy and low confidence.}

\label{fig:gradcam}
\end{figure}

Figure~\ref{fig:gradcam} presents representative Grad-CAM++ visualizations under different prediction scenarios. The selected examples include correctly classified high-confidence cases representing a common benign lesion (nv) (confidence: 0.988, entropy: 0.081), a malignant lesion (mel) (confidence: 0.941, entropy: 0.254), and a minority-class lesion (akiec) (confidence: 0.930, entropy: 0.319). In addition, an uncertain prediction case that was correctly classified (confidence: 0.553, entropy: 1.039) and identified by the uncertainty estimation module is presented. These visual explanations illustrate the image regions contributing to the model's classification decisions, while the uncertainty estimation mechanism helps identify cases requiring further expert assessment.

\section{Performance Evaluation}
\label{evaluation}

\subsection{Implementation Details}

The proposed framework is implemented using PyTorch and trained on a GPU-enabled environment. The models are optimized using the AdamW optimizer with a learning rate of $10^{-4}$ and a weight decay of $10^{-4}$. A cosine annealing learning rate scheduler is employed throughout training. Each model is trained for 15 epochs with a batch size of 16. Automatic mixed precision (AMP) training is employed to accelerate training, reduce GPU memory consumption, and maintain numerical stability. Gradient clipping with a maximum norm of 1.0 is applied to improve optimization stability.

\subsection{Experimental Results}

The classification performance of the proposed uncertainty-aware explainable framework is evaluated on the HAM10000 test dataset after applying the uncertainty-aware filtering criterion. The class-wise performance is summarized in Table~\ref{tab:classification_report}. Overall, the proposed framework achieves 96\% accuracy and a ROC-AUC of 0.99, demonstrating excellent discriminative capability for multi-class skin lesion classification.

\begin{table}[!ht]
\caption{Classification performance on the uncertainty-filtered test set.}
\label{tab:classification_report}
\centering
\begin{tabular}{lcccc}
\toprule
\textbf{Class} & \textbf{Precision} & \textbf{Recall} & 
\textbf{F1-score} & \textbf{Samples} \\
\midrule 
bkl   & 0.96 & 0.94 & 0.95 & 141  \\
nv    & 0.99 & 0.97 & 0.98 & 1044 \\
df    & 1.00 & 0.94 & 0.97 & 16   \\
mel   & 0.81 & 0.92 & 0.86 & 150  \\
vasc  & 1.00 & 0.96 & 0.98 & 26   \\
bcc   & 0.95 & 0.98 & 0.96 & 84   \\
akiec & 0.90 & 0.94 & 0.92 & 47   \\
\midrule
Accuracy & -- & -- & 0.96 & 1508 \\
ROC-AUC & -- & -- & 0.99 & 1508 \\
Macro Avg & 0.94 & 0.95 & 0.95 & 1508 \\
Weighted Avg & 0.96 & 0.96 & 0.96 & 1508 \\
\bottomrule
\end{tabular}
\end{table}

The proposed framework consistently achieves high precision, recall, and F1-score across most lesion categories. In particular, bkl, nv, df, vasc, and bcc exhibit F1-scores ranging from 0.95 to 0.98, indicating robust classification performance for both common and relatively underrepresented lesion classes. Although mel remains the most challenging category because of its high visual similarity to other pigmented lesions and substantial intra-class variability, the framework still achieves a recall of 0.86, which is clinically important for reducing missed melanoma diagnoses.

Overall, the framework obtains macro-average precision, recall, and F1-score of 0.94, 0.95, and 0.95, respectively, together with weighted-average scores of 0.96 across all three metrics. These results demonstrate that the proposed uncertainty-aware deep ensemble effectively addresses class imbalance while maintaining reliable and consistent performance across the seven diagnostic categories. The high ROC-AUC further confirms the framework's strong ability to distinguish between different skin lesion types, highlighting its potential for reliable computer-aided skin lesion classification.

\subsection{Confusion Matrix Analysis}

\begin{figure}
\centering
\includegraphics[width=\linewidth]{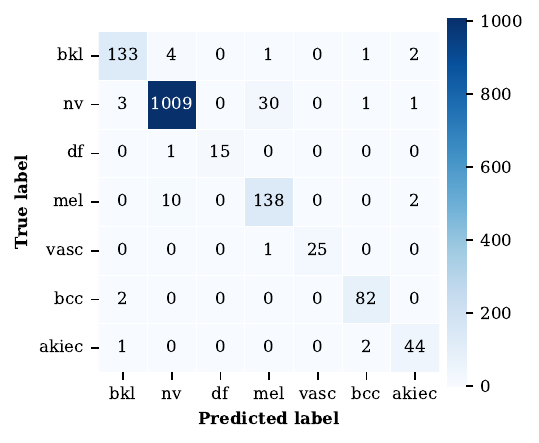}
\caption{Confusion matrix of the proposed framework on the uncertainty-filtered test set.}
\label{fig:cm}
\end{figure}

Figure~\ref{fig:cm} illustrates the confusion matrix of the proposed framework on the uncertainty-filtered test set. The confusion matrix provides a detailed visualization of the classification outcomes by showing the distribution of correct and incorrect predictions across the seven skin lesion categories.

Most predictions are concentrated along the main diagonal, indicating that the proposed framework correctly classifies the majority of skin lesion images. Only a small number of samples are misclassified across different lesion categories. The largest confusion occurs between mel and nv, where 10 mel samples are predicted as nv and 30 nv samples are predicted as mel. This observation is expected because these lesion types often exhibit similar dermoscopic characteristics, including pigmentation patterns, color variation, and lesion boundaries.

Minor misclassifications are also observed among bkl, bcc, and akiec, reflecting the visual similarities between these lesion categories. In contrast, df and vasc show very few classification errors, indicating that their distinctive visual characteristics are effectively captured by the proposed framework. Overall, the confusion matrix demonstrates that most errors occur between clinically similar lesion types, suggesting that the uncertainty-aware deep ensemble learns discriminative lesion representations while maintaining robust classification performance across all seven categories.

\section{Clinical Deployment Scenarios}
\label{clinical}

To demonstrate the potential practical applicability of the proposed uncertainty-aware framework, two clinical deployment scenarios are considered: large-scale skin lesion assessment with high prediction coverage and point-of-care decision support with high prediction confidence. The uncertainty estimation mechanism enables the framework to balance prediction coverage and reliability according to different clinical requirements.

\subsection{Large-Scale Skin Lesion Assessment (High Coverage)}

In large-scale skin lesion assessment, the objective is to classify a large number of dermoscopic images while maximizing prediction coverage. This scenario can be supported by applying a relatively relaxed uncertainty criterion, allowing the framework to provide predictions for a larger number of cases. However, accepting predictions with higher uncertainty may increase the likelihood of misclassification, particularly among visually similar lesion categories. Therefore, uncertain cases can be prioritized for dermatologist review, while the framework efficiently handles a large number of cases with high-confidence predictions.

\subsection{Point-of-Care Decision (High-Confidence Prediction)}

In point-of-care clinical decision support, prediction reliability is prioritized over coverage. Therefore, a stricter uncertainty filtering criterion is applied to retain only highly reliable predictions while rejecting uncertain cases for further assessment. In this work, predictions satisfying an entropy threshold of $\mathcal{H}(p)<1.0$ and a confidence threshold of $\max(p_c)\geq0.7$ are considered reliable. Although this strategy reduces prediction coverage by sending more cases for manual dermatologist review, it improves the reliability of automated predictions and provides more trustworthy decision support for individual clinical assessments.

\section{Conclusion}
\label{conclu}

This paper presents an uncertainty-aware and explainable deep learning framework for multi-class skin lesion classification using dermoscopic images. The proposed framework integrates advanced CNN and transformer-based architectures with deep ensemble learning and MC Dropout to enhance classification robustness and quantify predictive uncertainty, while Grad-CAM++ provides visual explanations to improve model transparency. Experimental evaluation on the HAM10000 dataset demonstrates that the framework achieves competitive classification performance with reliable uncertainty estimation and interpretable predictions. Under the uncertainty-aware filtering criterion (entropy $<1.0$ and confidence $\geq0.7$), the framework achieves 96\% accuracy and 99\% ROC-AUC, with competitive macro and weighted-average precision, recall, and F1-score performance. Despite these promising results, further validation on larger and more diverse multi-center datasets is required to assess generalizability across different clinical scenarios. Future work will focus on improving computational efficiency, exploring multimodal learning approaches, and developing more robust uncertainty estimation strategies for trustworthy AI-assisted skin lesion diagnosis.

\bibliographystyle{IEEEtran}
\bibliography{sn-bibliography}

\end{document}